# Higher-order photon correlation as a tool to study exciton dynamics in quasi-2D nanoplatelets

Daniel Amgar[1,†], Gaoling Yang[1,2,†], Ron Tenne[1], Dan Oron[1,*]

[1] Department of Physics of Complex Systems, Weizmann Institute of Science, Rehovot 76100, Israel

[2] Applied Nanophotonics Laboratory, Shenzhen MSU-BIT University, Shenzhen, Guangdong 518172, China

ABSTRACT

Colloidal semiconductor nanoplatelets, in which carriers are strongly confined only along one dimension, present fundamentally different excitonic properties than quantum dots, which support strong confinement in all three dimensions. In particular, multiple excitons strongly confined in just one dimension are free to re-arrange in the lateral plane, reducing the probability for multi-body collisions. Thus, while simultaneous multiple photon emission is typically quenched in quantum dots, in nanoplatelets its probability can be tuned according to size and shape. Here, we focus on analyzing multi-exciton dynamics in individual CdSe/CdS nanoplatelets of various sizes through the measurement of second-, third-, and fourth-order photon correlations. Thus, for the first time, we can directly probe the dynamics of the two, three and four exciton states in the single nanocrystal level. Remarkably, although higher orders of correlation vary substantially among the synthesis' products, they strongly correlate with the value of second order antibunching. The scaling of the higher order moments with the degree of antibunching presents a small yet clear deviation from the accepted model of Auger recombination through binary collisions. Such a deviation suggests that many-body contributions are present already at the level of triexcitons. These findings highlight the benefit of high-order photon correlation spectroscopy as a technique to study multi-exciton dynamics in colloidal semiconductor nanocrystals.

INTRODUCTION

Nanoplatelets (NPLs), colloidally synthesized two-dimensional nanocrystals (NCs), have shown great potential for low gain threshold lasing[1], [2], light emitting diodes[3], [4] and as photovoltaic sensitizers[5] due to their unique features: tunable band-gap, giant oscillator strength, narrow-band emission, and high lateral carrier mobility.[6] These superb properties arise due to a distinct exciton dynamics compared with their 0D and 1D counterparts. One important example is the recombination pathways of multi-excited states. When three or more charge carrier occupy a nanocrystal, an additional non-radiative recombination path opens up – the Auger process. In Auger recombination, an electron-hole pair recombines, and the excess energy is transferred to a third spectator charge. In quantum dots (QDs), the Auger rate for a biexciton (and higher multi-excited states) is much higher than the radiative recombination rate, significantly reducing the quantum yield (QY) of such states.[7] In contrast, due to the mean value of the lateral separation of excitons in a NPL, Auger rates are substantially lower than in QDs and the biexciton QY (BXQY) can approach that of the single exciton state.[8]

Previous studies concluded that since the electron-hole binding energy in NPLs is much higher than the lateral confinement energy, Auger recombination thus occurs through exciton-exciton collisions.[9] Using ensemble experimental approaches, such as time-resolved photoluminescence (PL) and transient absorption, Li *et al.* concluded that the rate of Auger relaxation decreases linearly with increasing the NPL lateral area and is inversely proportional to $d^7$, where d is the thickness.[10] Thus, biexcitons in large NPLs will preferably undergo radiative recombination whereas the Auger mechanism is the probable relaxation route for biexcitons in small NPLs (see figure 1a).[10]–[13] While these experiments were paramount to the understanding of multi-exciton dynamics in NPLs, they rely on measurements at high excitation powers, promoting the effect of charging and photo-bleaching, which may skew the conclusions. Moreover, as with all ensemble measurements, their interpretation is challenging due to dispersity in properties such as the absorption cross-section and the recombination rates. A different approach to probe multi-exciton dynamics is to measure second-order photon correlations in the PL of NCs.[14]–[16] While fluorescence correlation spectroscopy (FCS) is a common method to apply correlations in fluorescent light for biological imaging[17] and molecular spectroscopy,[18] typically, the observed timescales are beyond a microsecond. In contrast, photon correlations at the excited state lifetime scale are seldom applied for spectroscopy of molecules and nanostructures[19]–[21] and for microscopy applications[22]–[24]. However, such a method is naturally suitable to investigate multi-excited states since it observes the statistics of photon pairs emitted within a short delay.[8], [25] In a photon correlation measurement, light emitted from a single nanocrystal is split into two or more detectors – a Hanbury-Brown and Twiss setup (see figure 1b). When photon pairs are binned according to the delay time between detections a distinct dip at $\tau = 0$ is indicative of photon antibunching; a reduced probability for the detection of two photons simultaneously (see figure 1c and d). In colloidal nanocrystals, antibunching is a direct result of the Auger non-radiative decay pathway for biexcitons.[26] In this experiment, even at excitation intensities below saturation, one can isolate the rare events in which two excitons were simultaneously present and extract the rate of Auger recombination.[27] In a typical QD, the low BXQY leads to nearly complete antibunching and the second-order correlation function ($g^{(2)}(\tau)$) approaches vanishingly low probabilities at zero delay times (figure 1c). However, the plethora of nanocrystal structures produced in colloidal synthesis enables altering this property, for example by growth of larger QDs or graded shells.[28]–[31] While generally such a modification requires some relaxation of quantum confinement, growing large area NPLs enables the production of high multi-exciton QY particles without sacrificing the longitudinal confinement.[11] Using single particle spectroscopy, Ma *et al.* measured BXQY as high as 0.9 of the single exciton QY for such NPLs with a large lateral area.[8]

Here, we study the relations between photon correlation of orders two to four as a direct spectroscopic method to investigate the underlying mechanisms of multi-exciton recombination in NPLs. Surprisingly, while NPLs within a single sample span almost the entire range of second-order antibunching values (0-1), we found that the values of third and fourth order antibunching strongly correlate with that of the second order antibunching value. In addition, we show that while the well-accepted binary exciton collisions model captures a significant part of the dynamics, there are small yet significant deviations from it, indicating the effect of many-body interactions; we postulate that this modification is due to the tendency of a spectator exciton to Coulombically attract other excitons and thus promote Auger interactions.

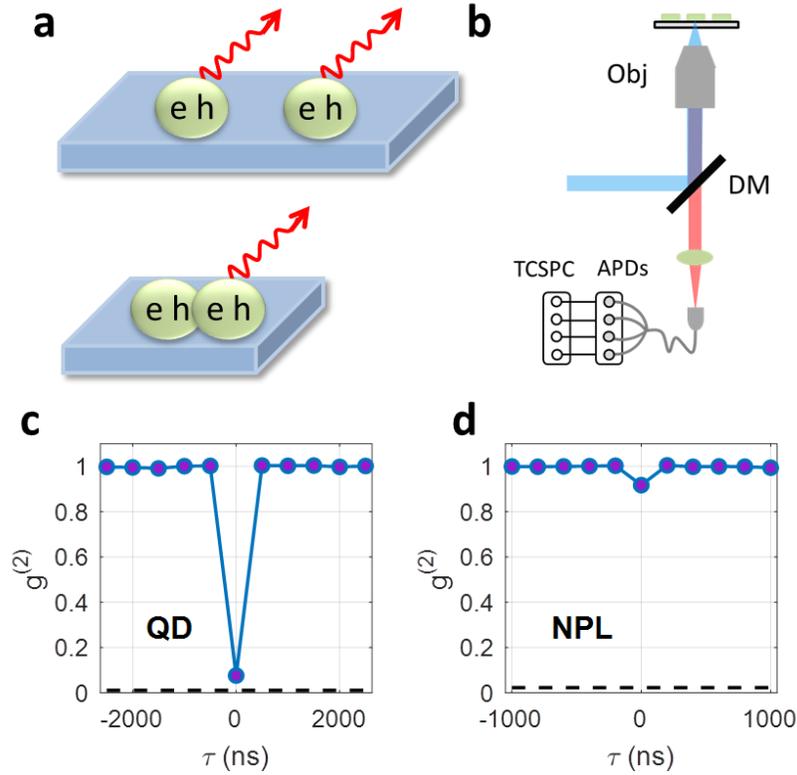

**Figure 1.** (a) Schematic illustration of the lateral size-dependent Auger effect in NPLs showing that in large NPLs, spatially separated excitons can lead to multi-exciton emission while in small NPLs close excitons can lead to enhanced Auger process and only one emitted photon. (b) Single-particle optical setup. Obj is an objective, DM is a dichroic mirror, APD is an avalanche photodiode, and TCSPC is time-correlated single-photon counter. Typical antibunching dip of a single quantum dot (c) compared with a single NPL (d), showing their different tendency to emit multi-excitons. The black dashed line indicates the estimated background level due to detector dark current.

RESULTS AND DISCUSSION

Cores of colloidal 5 monolayers CdSe NPLs were synthesized according to a previously reported procedure.[11] By varying the reaction time we fabricated NPLs batches with three different average lateral sizes: ~5x12 nm, ~9x32 nm, and ~14x41 nm; in the following, these samples are referred to as small, medium, and large area NPLs, respectively. Further details about the synthesis can be found in the Materials and Methods section. Figure 2a-c presents transmission electron microscope (TEM) images of the small, medium, and large CdSe samples. Figure S1 in the Supporting Information (SI) presents the lateral size distributions as analyzed from TEM images. Absorption and PL spectra for the medium NPLs sample are shown in figure 2d. All three samples presented very similar spectra with slightly shifted peak positions as presented in figure 2e. The lowest energy excitonic peaks (electron-heavy hole) are centered at ~542, ~551, and ~549 nm, and the PL peaks at ~547, ~555, and ~553 nm for small, medium, and large area NPLs, respectively.

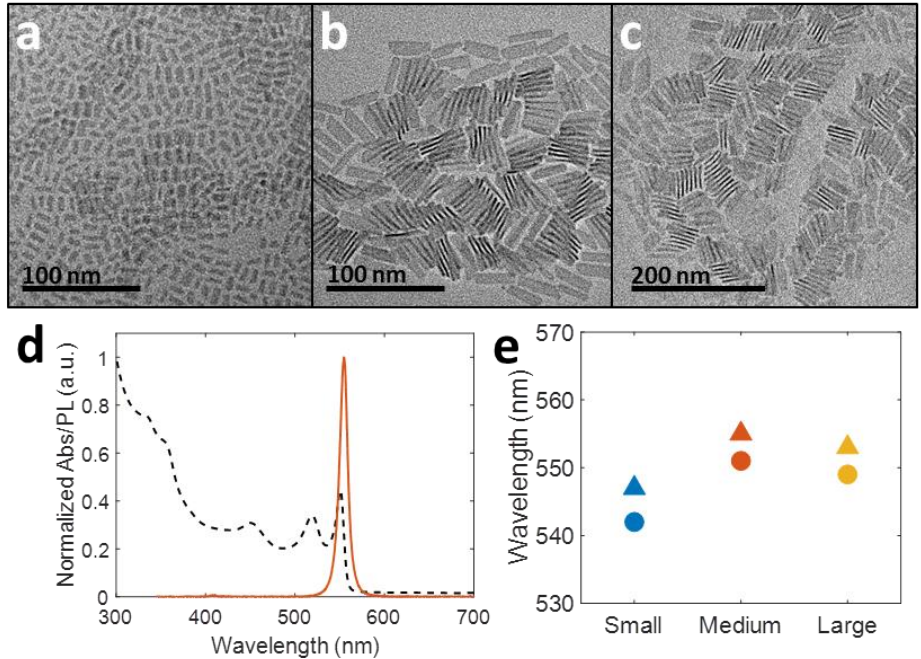

**Figure 2.** TEM images of small (a), medium (b), and large (c) NPLs. (d) Absorption and PL spectra of the medium NPLs. (e) Absorption (circles) and PL (triangles) peak positions for small (blue), medium (orange), and large (yellow) NPLs, as indicated in the x-axis.

Since photon correlation measurements require a high QY and long-term photo-stability under relatively high excitation intensities, we passivated the surface of the NPLs cores by growing CdS shells of 3 monolayers.[9], [32], [33] TEM images, absorption, and PL spectra of the core/shell NPLs are shown in the SI, figure S2. Although the growth of the CdS shell slightly altered the lateral dimensions of the NPLs, probably due to CdS growth on the edges of the NPLs, we assume that this has a negligible effect on our results due to the inherent size dispersity of our samples and the use of single particle spectroscopy.

To understand the role of lateral size on the multi-exciton interactions we use a single-particle PL characterization setup to measure correlations in the emitted photon stream. The setup, schematically shown in figure 1b, is comprised of a standard confocal microscope with a pulsed laser excitation (470 nm, ~100 ps pulses) focused onto the sample plane by an objective lens. Light emitted by the NPL is collected *via* the same objective lens and imaged on a multi-mode fiber splitter, dividing the PL equally among four single photon avalanche detectors (SPADs). The use of four SPADs enables the measurement of short-time second-, third-, and fourth-order photon correlations.[34] The photons' detection times are clocked and digitally stored by a time-correlated single-photon counting (TCSPC) module (Hydra-harp 400, PicoQuant). Additional details on the single particle spectroscopy setup are given in the Materials and Methods section.

Figure 3a depicts the second-order correlation as a function of delay time ( $G^{(2)}(\tau)$ ) of a representative NPL from the medium-sized sample, presenting the number of detected photon-pairs *versus* the delay time between detections at the resolution of the laser pulse repetition period. In order to isolate the effect of antibunching on the correlation functions from that of classical fluctuations, we assign unity value to the plateau at non-zero, yet relatively short, delay times. Thus, we define $g^{(2)}(\tau) \equiv \frac{G^{(2)}(\tau)}{G^{(2)}(plateau)}$, where in our analysis $G^{(2)}(plateau)$ refers to the average value between one and five pulse delays. With this definition, $g^{(2)}(0)$ is an estimate for antibunching in the bright 'on' state of the nanocrystal. For the measurement shown in figure 3a, this value is $g^{(2)}(0) = 0.7677 \pm 0.0005$.

The high photon emission rate and BXQY together with the minutes-long photo-stability of core/shell NPLs enable us to go beyond the standard measurements of antibunching and measure third- and fourth-order photon correlations from a single colloidal NC. Figure 3b presents the third-order correlation ( $G^{(3)}(\tau_1, \tau_2)$ ) results, analyzed from the same data set used in figure 3a, a measure of the number of detected photon triplets *versus* the delay times between the three photons. To construct $G^{(3)}$ we randomly assign the numbers 1-3 to each detected triplet and calculate the delay time $\tau_1 = t_2 - t_1$ and $\tau_2 = t_3 - t_1$ , where $t_i$ is the detection time of photon $i$ (see SI for more details on the analysis) . The vertical, $(0, \tau_2)$, horizontal, $(\tau_1, 0)$, and diagonal, $(\tau_1 = \tau_2)$, lines (except for the point $(0,0)$) signify the detection of two photons at the same time and a third delayed photon. The rest of the points describe three photons detected at different times. Most importantly, the probability to detect three photons at the same time is depicted in the point of origin, where $\tau_1 = \tau_2 = 0$, presenting the lowest value. From the ratio of this value to the plateau we calculate $g^{(3)}(0,0) = 0.46 \pm 0.004$.

As a complementary analysis for multi-exciton dynamics, we observe the detections' delay time relative to the exciting laser pulse. Each photon triplet arriving after the same excitation pulse is split into the first, second, and third arriving photons. We can then generate a separate lifetime curve for the triexciton, biexciton, and single exciton states, respectively (SI, section S1). Analyzing the PL of another single NPL (not the one shown in figure 3), the individual lifetime curves were fit with a bi-exponential function, from which we extract the effective recombination decay rates for the different states ($\tau_{1x} = 6.7\ ns$, $\tau_{2x} = 2.6\ ns$, $\tau_{3x} = 1.0\ ns$). Using these rates, we estimate the values of $g^{(2)}(0)$ and $g^{(3)}(0,0)$ to be 0.77 and 0.35, respectively, for the analyzed NPL (details are found in SI, section S1). Calculated values only roughly agree with the measured correlation values, $g^{(2)}(0) = 0.81$ and $g^{(3)}(0,0) = 0.52$, presumably since the model is limited to the case of a single exponential model while the data clearly exhibits more complex dynamics.

For the brightest of NPLs, one can go another step further and analyze the detection of four simultaneous photons. The fourth-order correlation function ( $G^{(4)}(\tau_1, \tau_2, \tau_3)$ ) results for the same NPL shown in figure 3a,b are extracted from the same measurement to produce the images presented in figures 3c,d. Since $G^{(4)}$ is a 3-variable function of the three delay times ($\tau_1, \tau_2, \tau_3$), it is more challenging to visualize. Thus, for clarity we present two constant $\tau_3$ cross-sections: at zero ( $G^{(4)}(\tau_1, \tau_2, 0)$ ) and one ( $G^{(4)}(\tau_1, \tau_2, 1)$ ) pulse delay time. The probability to detect four photons emitted simultaneously, derived from the center point in figure 3c, is $g^{(4)}(0,0,0) = 0.25 \pm 0.04$. As expected, $g^{(3)}(0,0)$ and $g^{(4)}(0,0,0)$ values are smaller than $g^{(2)}(0)$ because the multi-exciton non-radiative recombination rates increase with the number of excitons. Naively, the Auger recombination rate follows the number of exciton-pair permutations in the state while the radiative rate grows linearly with number of excitons. This results in a lower QY for higher orders of multi-excitations. A more comprehensive explanation of the higher-order correlation analyses and examples of $G^{(2)}$-$G^{(4)}$ plots of single NPL from the small-sized and large-sized samples are shown in SI section S2 and figure S4.

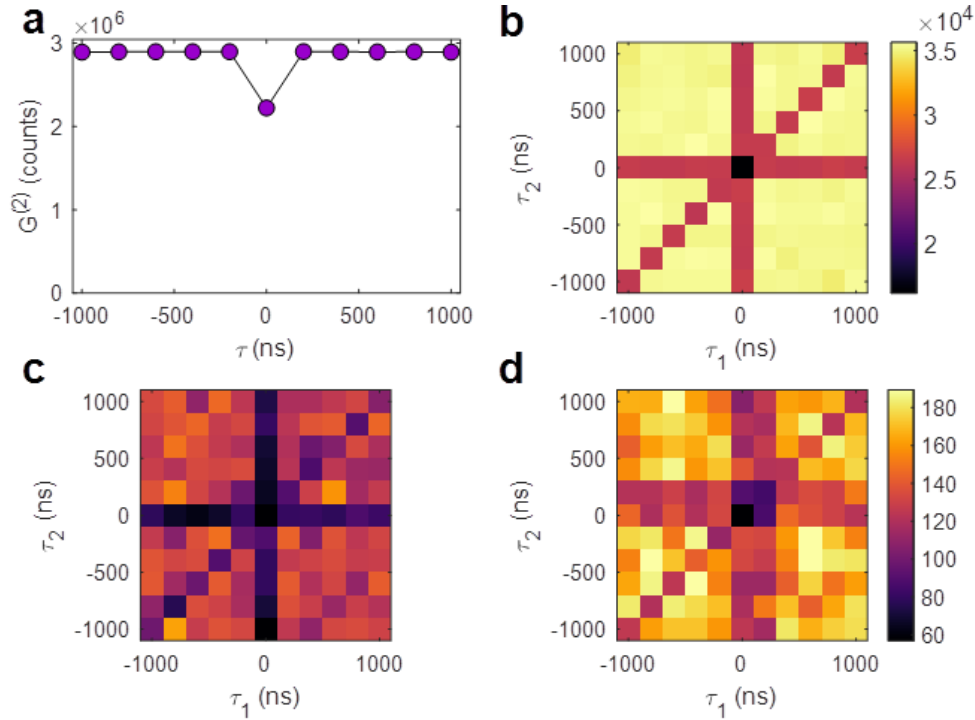

**Figure 3.** Photon correlations of a representative single NPL from the medium area sample. (a) Second-order antibunching [$G^{(2)}(\tau)$]. (b) Third-order antibunching [$G^{(3)}(\tau_1, \tau_2)$]. (c,d) 2D cross-section of the fourth-order correlation function [$G^{(4)}(\tau_1, \tau_2, \tau_3)$] at (c) zero delay time ($\tau_3 = 0$) and(d) 1 pulse delay time ($\tau_3 = 1$ pulse). The experimental parameters of the measurements: excitation power of the laser: ~90 nW; repetition rate of the laser: 5 MHz; measurement acquisition time: 180 s.

To some extent, photon correlations can be affected by the average number of excitations per nanocrystal,[27] especially when approaching the saturation intensity. In order to confirm that we work with below-saturation excitation powers, we performed a saturation experiment (described in section S3 in the SI).[35] Figure S5, presenting the PL intensity for NPLs *versus* excitation power, demonstrates that saturation does not occur even at the highest pulse energy of $\sim 9 \cdot 10^{-14}$ J per pulse for the medium area sample. In order to ensure below-saturation excitation powers for all NPLs samples we use only $\sim 1.8 \cdot 10^{-14}$ J per pulse. An estimation for the average number of absorbed photons per pulse, calculated according to absorption cross-section estimated in literature and described in full in the SI section S4, yields an average of 0.04, 0.2 and 0.4 excitons per pulse per NC for the small, medium, and large area samples, respectively.[8],[36], [37] While the average population for the larger samples approaches saturation, the biexciton population is still substantially smaller than that of the single excitons. Therefore, it should only slightly affect the correlation function measurements.[27] In order to allow complete relaxation of excitons between laser pulses, the repetition rate was set to 5MHz (200 ns between subsequent pulses), much longer than the exciton lifetime ($\sim 7$ ns).

Unlike for the case of single QD spectroscopy, where emission of photon pairs is strongly suppressed and a common metric for identification of single emitters is $g^{(2)}(0) < 0.5$, the antibunching dip magnitude alone cannot be used as a signature of measuring a single NPL. We therefore use several different steps to ensure that our measurements are not contaminated with results from NPL clusters. First, we prepare sparse samples in which bright spots are separated by $\sim 5$ $\mu m$ on average. Second, by observing fluorescence intermittency ("blinking") we exclude measurements which do not present repeated periods of background-level brightness (see figure S6 in the SI). Finally, we apply a time-gating test for the $g^{(2)}$ function of each measurement. Photons that arrive at short delays relative to the excitation pulse are filtered-out and only late-arriving photons that originate preferentially from single excitons are used to construct the $g^{(2)}(\tau)$ curve.[8], [38] Then, by plotting the calculated $g^{(2)}(0)$ values *versus* increasing gating times, beyond the biexciton lifetime, we expect that single NPLs would show a significant decrease in $g^{(2)}(0)$. Measurements whose $g^{(2)}(0)$ falls below 0.5 after this filtering procedure are considered single NPLs and used for further analysis (the process is demonstrated in figure S7 in the SI). This 0.5 threshold was selected based on the formula $g^{(2)}(0) = 1 - \frac{1}{n}$, where $n$ is the number of emitters.[39] From over 200 performed measurements of all three samples, 151 met the single particle criterion and presented a signal-to-noise ratio (SNR) of more than 10 for $g^{(3)}(0,0)$. The NPLs exhibit a broad distribution of $g^{(2)}(0)$ and $g^{(3)}(0,0)$ values, as shown in the histograms in figure S8 and table S1of the SI.

Figure 4a presents the dependence of the $g^{(3)}(0,0)$ values of all the measured NPLs on $\left[g^{(2)}(0)\right]^2$. Surprisingly, all the measurements follow a universal behavior, lying on a distinct monotonic line despite the very large variance of both $g^{(2)}(0)$ and $g^{(3)}(0,0)$. We attribute this observation to the fact that the QY of multi-excitons is dependent on the Auger recombination rate which is determined by the NPL's lateral size. Therefore, while our synthesis products vary in aspect ratio and transverse size, the BXQY and triexciton QY are both essentially dependent on the NPL's area. In accordance with this principle the average $g^{(2)}(0)$ for small area NPLs (blue circles), is lower compared with that of the medium area NPLs. Large area NPLs, present the highest $g^{(2)}(0)$ values from the three samples. This finding confirms the aforementioned trend of size-dependent antibunching in NPLs.[10]˙[8]

Note that while a qualitatively similar trend of third *versus* second order antibunching has been observed due to the addition of a Poissonian background to the fluorescence of a single photon emitter[15], the deep blinking contrast in our measurements ensures us that this is not the case here. In a typical measurement, the 'on' state PL rate is more than 50 times higher than that of the 'off' state and thus fluorescent background accounts for less than 2% of the detected photons (see figure S6, SI).

In order to examine this remarkable correspondence of second- and third-order correlations, we attempt to compare this dependence to an exciton-exciton collision model without any fit parameters (black solid line in figure 4a). The simplified kinetic model describes exciton-exciton interactions, assuming that electron-hole pairs in quantum wells are tightly-bound and thus Auger recombination follows second-order kinetics, *i.e.* it requires a collision of two excitons.[9], [40] A detailed description of the model is found in SI section S5 and in references 9 and 10. In short, the second-order correlation function at zero delay time can be expressed as:

$$g^{(2)}_{model}(0) = \frac{2k_{rad}}{2k_{rad} + k_{Aug}} \tag{1}$$

where $k_{rad}$ is the radiative decay rate of a single exciton and $k_{Aug}$ is the Auger recombination rate of the biexciton state. Following a similar logic, the third-order correlation function can be expressed as a function of $g^{(2)}(0)$:

$$g^{(3)}_{model}(0,0) = \frac{3k_{rad}}{3k_{rad} + 3k_{Aug}} \cdot g^{(2)}(0) = \frac{\left[g^{(2)}(0)\right]^2}{2 - g^{(2)}(0)}, \tag{2}$$

where we consider all possible exciton combinations, $\binom{3}{2} = 3$, for the Auger process. We note that the non-classical nature of photon statistics can also be examined in the third-order photon correlation using the inequality $g^{(3)} < \left[g^{(2)}\right]^2$.[21], [41] Indeed, the expression in equation 2 fulfills the inequality for all values of $g^{(2)}$ smaller than unity, in agreement with the standard antibunching criterion.

A careful look at the model with respect to the experimental data in figure 4a reveals that the experimental $g^{(3)}(0,0)$ values are consistently smaller than the model's expectation. Figure 4b highlights the deviation of our results from the above mentioned model (Eq. 2), presenting the difference between the two for each measurement point. To supply some quantitative estimate of this deviation without precise knowledge of the underlying model, we average these differences in two regions of this graph. For low $\left[g^{(2)}(0)\right]^2$ (32 measurements between 0 and 0.3) the weighted average is only 2.6 standard deviations below the model, showing a clear tendency towards values lower than predicted by the model. Even more significantly, the average difference at higher $\left[g^{(2)}(0)\right]^2$ values (96 measurements between 0.3 and 0.8) is more than 13 standard deviations below zero. This deviation shows a statistically significant disagreement between our results and the biexciton collision model and a trend of greater deviation with larger $g^{(2)}(0)$ values.

The downward deviation for moderately antibunched particles suggests that these NPLs exhibit a higher triexciton Auger recombination probability than the sum of Auger probabilities for all possible exciton pairs. We speculate that the significant triexction interaction term is the result of an enhanced Coulomb interaction (low dielectric constant environment) between excitons in NPLs. Such enhanced interaction, manifesting in the high exciton and biexciton binding energies of NPLs, can reduce the average exciton-exciton distance in the presence of a third exciton.[1], [11]

In order to better understand the magnitude of the multi-body effect, the results were fit to a phenomenological model that includes another contribution to the non-radiative decay rate of a triexciton beyond random exciton-exciton collisions: $k_{3B}$ (3B = three body).

$$k_{3x} = 3k_{rad} + 3k_{Aug} + k_{3B} \tag{3}$$

The resulting corrected model for the photon correlation function is presented in full in SI section S6. The dashed line in figure 4b represents a fit of the phenomenological model to the experimental results, yielding $k_{3B} \cong (0.12 \pm 0.03) \cdot k_{Aug}$. This value can be interpreted as a ~4% enhancement of the exciton-exciton Auger recombination rate in the presence of an additional exciton in the NPL.

Figure 4c presents the dependence of the $g^{(4)}(0,0,0)$ values of all the measured NPLs against $\left[g^{(2)}(0)\right]^3$ along with the prediction of the exciton-exciton collision model. For clarity we present here only the 29 measurements for which the SNR of $g^{(4)}(0,0,0)$ is more than 4. The analysis of $g^{(4)}(0,0,0)$ results is more challenging due to the low SNR of counting the rare events in which four simultaneous photons are detected. Nevertheless, the results indicate that, as in the case of triexcitons, four-exciton Auger rates are higher than predicted and marginally support the observed trend for the $g^{(3)}(0,0)$ results. Similarly to the $g^{(3)}(0,0)$ case, we applied a statistical test to quantify how much the $g^{(4)}(0,0,0)$ results deviate from the exciton-exciton collision model (the deviations are shown in figure 4d). As a result of low SNR in the case of $g^{(4)}(0,0,0)$, only the 21 measurements, which fall in the range between 0.3 and 0.8, were tested, yielding an average difference of 0.8 standard deviations below the model.

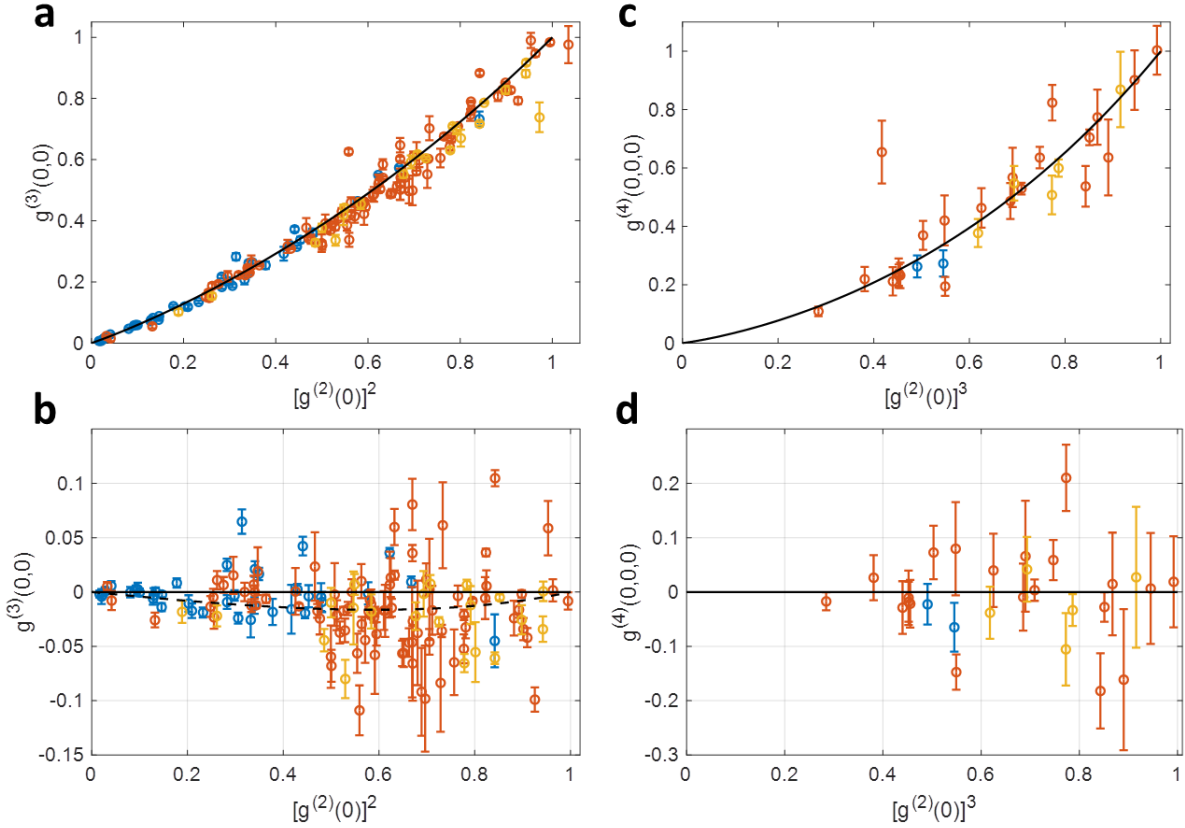

**Figure 4.** Summarized results for all measured individual NPLs, who had met the SNR and single-particle criteria.(a) Third order antibunching, $g^{(3)}(0,0)$, *versus* $\left[g^{(2)}(0)\right]^2$(b) Fourth order antibunching $g^{(4)}(0,0,0)$ *versus* $\left[g^{(2)}(0)\right]^3$. Black solid lines present $g^{(3)}(0,0)$ and $g^{(4)}(0,0,0)$ calculated with the binary collision model in (a) and (b) respectively. (c) Deviations of $g^{(3)}(0,0)$ from the binary collision model. The adjusted phenomenological model is shown in the black dashed line. (d) Deviations of $g^{(4)}(0,0,0)$ from the binary collision model. The colors blue, orange, and yellow correspond to measurement from the small, medium, and large area samples, respectively.

## CONCLUSIONS

In summary, we demonstrate the use of higher-order photon correlation measurements for spectroscopy, investigating the interaction between excitons in single CdSe/CdS core/shell NPLs. A comprehensive experimental study of two, three, and four simultaneous photon emission from a single NPL, shows that their probabilities are highly correlated. A careful glance at the scaling of $g^{(3)}(0,0)$ with respect to $g^{(2)}(0)$ reveals clear deviation from the well-accepted binary collision model indicating that many-body interactions play a significant role in the relaxation of multi-excitonic states. Our findings may affect the implementation of NPLs in light-emitting diode and lasers, where the QY of multi-exciton states are critical for high performance. In addition, the method and modeling used here can be applied to the study of multi-exciton states in different types of NCs that have non-zero BXQY and in particular NPLs from different material systems.

## MATERIALS AND METHODS

*Chemicals:* Cadmium nitrate tetrahydrate ($\geq$ 99.0%, Sigma), methanol ($\geq$ 99.95%, Bio-Lab), sodium myristate ($\geq$ 99%, Sigma), 1-octadecene (ODE, 90%, Sigma), Selenium (Se, $\geq$ 99.5%, Sigma), Cadmium acetate dehydrate ($\geq$ 98.0%, Sigma), oleic acid (OA, 90%, Sigma), ethanol (Gadot), N-methylformamide (NMF, 99%, Sigma), aqueous ammonium sulfide (40-48 wt. %, Sigma), hexane ($\geq$ 95%, Bio-Lab), acetonitrile ($\geq$ 99.97%, Bio-Lab), toluene ($\geq$ 99.7%, Bio-Lab), oleylamine (OLA, 80-90%, Sigma).

*Preparation of Cadmium Myristate precursor.* Cadmium nitrate (1.23 g) was dissolved in 40 ml of methanol. 3.13 g of sodium myristate were dissolved in 250 ml of methanol using strong stirring for one hour. After complete dissolution, the two solutions were mixed, resulting in a white precipitate. The precipitate was filtered and washed using a Buchner vacuum flask and dried under vacuum for 12 hours.[11]

*Synthesis of CdSe core nanoplatelets.* CdSe NPLs were synthesized according to a previous procedure from the literature with small modifications.[11] 170 mg of Cadmium myristate were dissolved in 15 ml octadecene (ODE) and degassed for 20 minutes. Then, temperature was raised to 240°C under Ar flow and a Selenium precursor solution (12 mg of Selenium in 1 ml ODE) was swiftly injected into the flask. One minute later, 80 mg of Cadmium acetate dehydrate were rapidly added into the flask and after 10 minutes the reaction was stopped and cooled down to room temperature. 1.5 ml of oleic acid (OA) were added at 210°C to stabilize the forming NCs. After the synthesis, the product was centrifuged with ethanol (1:1) at 6000 rpm for 5 minutes to get 5 ML-thick CdSe cores.

*Synthesis of CdSe/CdS core/shell nanoplatelets.* CdS shell growth was done according to Yang *et al.*[32]. To the washed CdSe cores, 1 ml of N-methylformamide (NMF) and 50 µl aqueous ammonium sulfide were added as a sulfur source to create phase transfer of the NPLs from hexane to NMF. After complete phase transfer the hexane was discarded and this step was repeated a second time. In order to avoid nucleation of CdS, excess $S^{2-}$ ions were removed from NMF as follows; 1.5 ml of acetonitrile and 1 ml toluene were added to precipitate the NPLs at 3800 rpm for 3 minutes. The precipitate was dispersed in 1 ml of NMF and the last step was repeated a second time with 1 ml of acetonitrile and 2 ml of toluene at 6000 rpm for 5 minutes. The precipitate was dispersed in 0.5 ml NMF and 1.5 ml of Cadmium acetate dehydrate in NMF solution (0.2 M) were added as a Cd source to further grow the shell under stirring for a few minutes. 4 ml of toluene were added to precipitate the NPLs and then dispersed in 1 ml NMF. Afterwards, 4 ml of hexane, 100 µl of oleic acid, and 100 µl of oleylamine were added under stirring for a few minutes till complete phase separation. The formed oleylamine-capped core/shell CdS/CdSe NPLs were collected. This cycle was repeated 3 times to produce 3 ML of CdS shell onto the CdSe cores.

*Single particle spectroscopy setup.* A 470 nm pulsed laser diode with maximal 20 MHz repetition rate (Edinburgh Instruments, EPL-470) was used for single particle excitation. The excitation laser was coupled into a microscope (Zeiss, Axiovert 200 inverted microscope) and focused using a high numerical aperture oil-immersion objective (Zeiss, Plan Apochromat X63

NA 1.4). The epi- detected signal was filtered, using a dichroic mirror (Semrock, Di02-R488-25 × 36) and a long-pass filter (Semrock, BLP01-488R-25), and coupled into a multimode fiber that equally splits the signal into four avalanche photodiode detectors (PerkinElmer, SPCM- AQ4C) that were connected to a time-correlated single-photon counting (TCSPC) system (Picoquant, HydraHarp 400). Single NPL saturation experiments were performed with the same setup by varying the laser excitation power in a triangular pattern (see more details in SI section S3)

*Characterization methods.* TEM images were taken on a JEOL 2100 TEM equipped with a LaB6 filament at an acceleration voltage of 200 kV on a Gatan US1000 CCD camera. UV-vis absorption spectra were measured using a UV-vis-NIR spectrometer (V- 670, JASCO). Fluorescence spectrum was measured using USB4000 Ocean Optics spectrometer excited by a fiber coupled 407 nm LED in an orthogonal collection setup.

ASSOCIATED CONTENT

**Supporting Information**.

Further EM and light spectroscopy of the NPLs and formulation of the models used within the manuscript. This material is available free of charge via the Internet at http://pubs.acs.org.

AUTHOR INFORMATION

**Corresponding Author**

[*]Dan.Oron@weizmann.ac.il

**Author Contributions**

[†]These authors contributed equally.

**Funding Sources**

Funding by the European Research Council Consolidator grant ColloQuantO and by the Crwon center of Photonics is gratefully acknowledged. DO is the incumbent of the Harry Weinrebe professorial Chair of Laser Physics.

**Notes**

The authors declare no competing financial interest.

ACKNOWLEDGMENT

The authors thank Maria Chekhova for helpful discussions.

# Supporting Information:

# Higher-order photon correlation as a tool to study exciton dynamics in quasi-2D nanoplatelets

*Daniel Amgar[1,†], Gaoling Yang[1,2,†], Ron Tenne[1], Dan Oron[1,\*]*

[1] Department of Physics of Complex Systems, Weizmann Institute of Science, Rehovot 76100, Israel

[2]Applied Nanophotonics Laboratory, Shenzhen MSU-BIT University, Shenzhen, Guangdong 518172, China

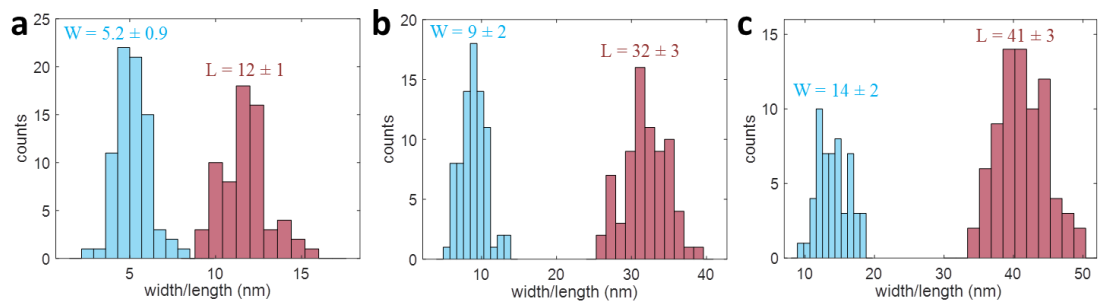

**Figure S1.** Size distributions of the width (W) and length (L) for small (a), medium (b), and large (c) CdSe core nanoplatelets. The size distributions were produced using ImageJ software.

# Figure S2: TEM and emission and absorption spectra for core/shell NPLs

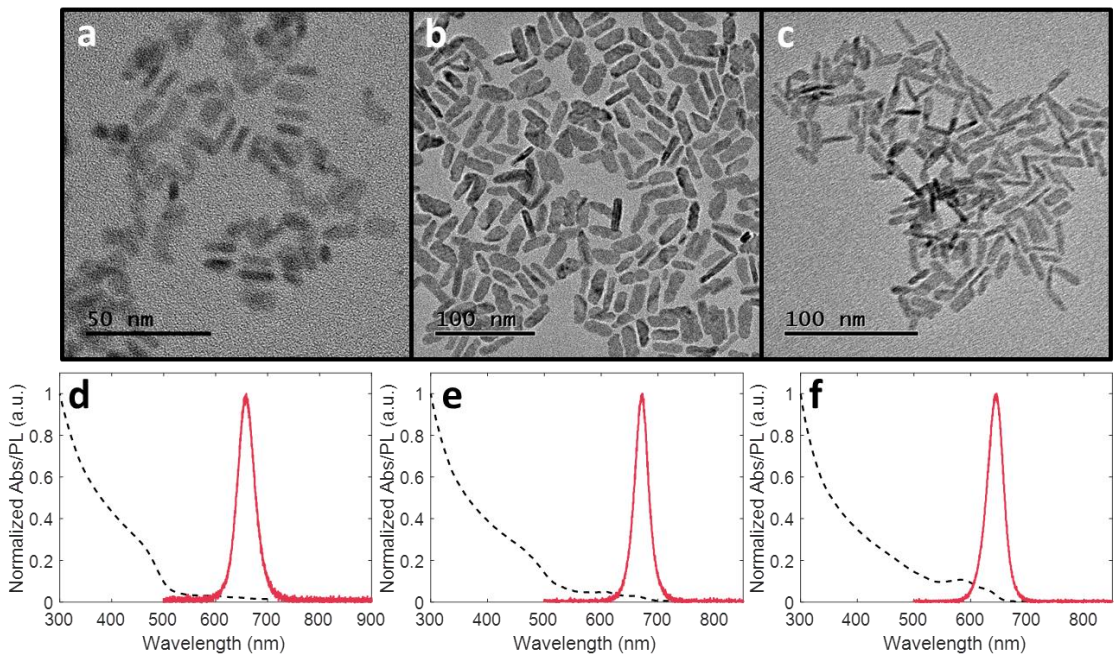

**Figure S2.** TEM images (a-c) and the corresponding absorption (black, dashed line) and photoluminescence (red line) spectra (d-f) of small (a,d), medium (b,e), and large (c,f) CdSe core/shell NPLs, respectively.

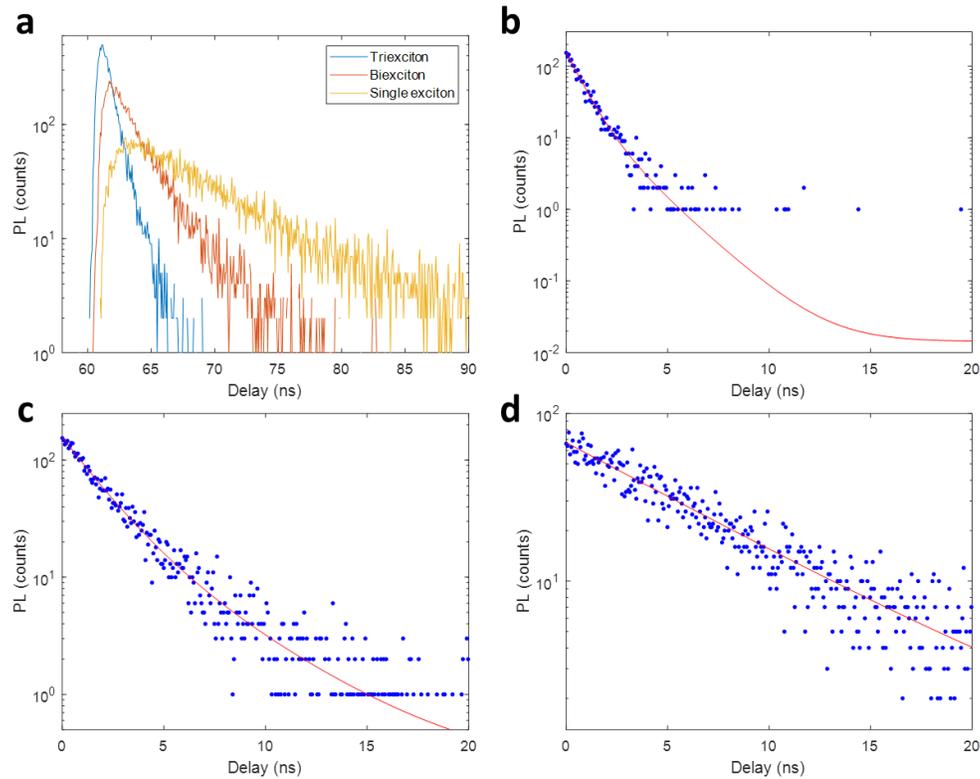

**Figure S3.** (a) Lifetime curves of three photons emitted together as a photon triplet from a representative single medium-sized NPL. (b-d) The individual lifetime curves of the triexciton, biexciton, and single exciton, respectively, with bi-exponential (b,c) and single exponential (d) fit.

Lifetimes of photons from the different multi-exciton were estimated from a single medium-sized NPL. First, we isolate photon triplets detected during the same excitation cycle. We then separate the photon into groups of first, second and third detected photons matching the relaxation of the tri-exciton, biexciton and single exciton respectively. Finally, we histogram each group separately according to the delay of detection relative to the laser pulse. The decay potion of each of the histograms was fit with a double exponential decay. From the exponential fits, shown in figures S3 (b-d) we obtain an effective lifetime for each state by performing a weighted average

$$\tau_{1x} = 6.7 \, ns, \; \tau_{2x} = 2.6 \, ns, \; \tau_{3x} = 1.0 \, ns \, , \qquad \text{(S1)}$$

where $\tau_{1x}$ is the lifetime of a single exciton (third photon), $\tau_{2x}$ is the lifetime of a biexciton (second photon), and $\tau_{3x}$ is the lifetime of a triexciton (first photon). As expected, $\tau_{3x}$ is the shortest lifetime since the triexciton recombination rate includes the largest number of pathways for radiative and non-radiative recombination. The biexciton state is longer-lived with a lifetime $\tau_{2x}$ and finally the single exciton state lifetime, $\tau_{1x}$, is the longest. Using the relation between lifetimes and recombination rates:

$$\frac{1}{\tau_{1x}} = k_{1x}^{rad} \qquad \text{(S2)}$$

$$\frac{1}{\tau_{2x}} = 2 \cdot k_{2x}^{rad} + \, k_{2x}^{Aug} \qquad \text{(S3)}$$

$$\frac{1}{\tau_{3x}} = 3 \cdot k_{3x}^{rad} + \, 3 \cdot k_{2x}^{Aug} \qquad \text{(S4)}$$

We estimate $g^{(2)}(0), g^{(3)}(0,0)$:

$$g^{(2)}(0) = \frac{2k_{1x}^{rad}}{2k_{1x}^{rad} + k_{2x}^{Aug}} = \frac{2\tau_{2x}}{\tau_{1x}} = 0.77 \qquad \text{(S5)}$$

$$g^{(3)}(0,0) = \frac{3k_{1x}^{rad}}{3k_{1x}^{rad} + 3k_{2x}^{Aug}} \cdot g^{(2)}(0) = \frac{3\tau_{3x}}{\tau_{1x}} g^{(2)}(0) = 0.35, \qquad \text{(S6)}$$

where $k_{1x}^{rad}$ is the radiative decay rate of a single exciton and $k_{2x}^{Aug}$ is the nonradiative Auger rate of a biexciton.

The values estimated directly from the correlation function measurements are $g^{(2)}(0) = 0.81$, $g^{(3)}(0,0) = 0.52$.

While the values from both estimates only roughly agree, we note that the calculation of $g^{(2)}(0)$ and $g^{(3)}(0,0)$ from the lifetime is only approximate since the above mentioned model assumes a single exponential decay. In contrast, figure S3 clearly shows that a bi-exponential function is necessary to fit the data.

## Supporting Information section S2: Higher-order correlation presentation

The third-order correlation function ($g^{(3)}(\tau_1, \tau_2)$) reflects the probability to detect photon triplets along delay times $\tau_1$ and $\tau_2$. To estimate this function we locate within our data detection triplets whose timings are within a set range of $5 \cdot t_{Pulse}$ ($t_{Pulse}$ is the time between consecutive laser pulses). To avoid artifact of detector dead time we use only triplets from three different detectors. In order to avoid data analysis artifacts, we randomly assign the integers 1,2 and 3 to each detection triplet. $\tau_1$ then represents the delay time between the arrival times of the first and second photons ( $\tau_1 = t_2 - t_1$ ) and $\tau_2$ represents the delay time between the arrival times of the first and third photons ( $\tau_2 = t_3 - t_1$ ), where $t_i$ are the detection times rounded to the laser trigger time preceding the i-th detection. We state that three photons are detected simultaneously if $\tau_1 = \tau_2 = 0$ - emitted within a certain pulse (time between pulses is 200 ns).

Following a similar logic, the $g^{(4)}$ function is presented as a function of the delay times among four detections, where $\tau_1$ is the delay time between the arrival times of the first and second photons ( $\tau_1 = t_2 - t_1$ ), $\tau_2$ is the delay time between the arrival times of the first and third photons ( $\tau_2 = t_3 - t_1$ ), and $\tau_3$ is the delay time between the arrival times of the first and fourth photons ( $\tau_3 = t_4 - t_1$ ). The color bar on the right in figure 3 (main text) represents the number of the 4-photon detections at each delay time.

Figure S4 demonstrates $g^{(2)}$, $g^{(3)}$, $g^{(4)}$ measurements of single NPL from the small-sized sample (a-d) and from the large-sized sample (e-h).

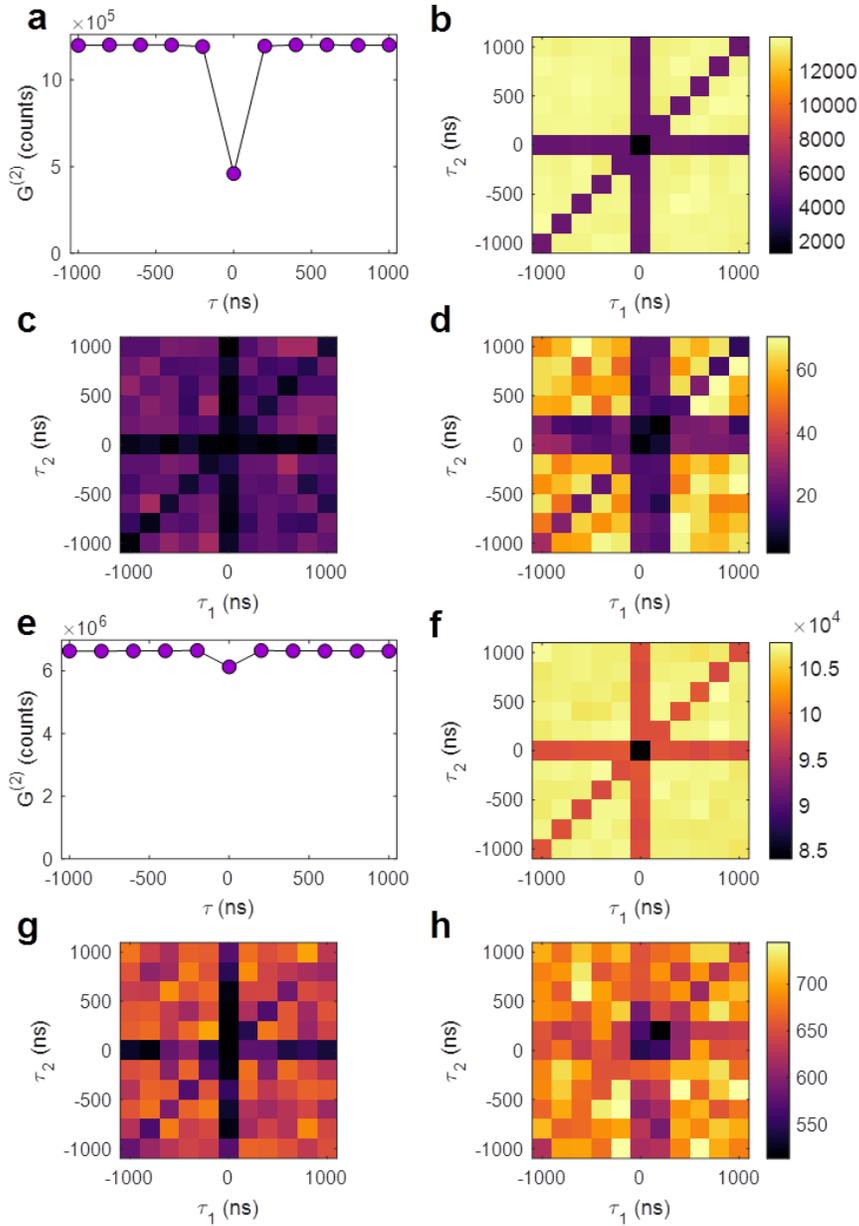

**Figure S4.** Second-order antibunching [$g^{(2)}(\tau)$] (a,e), third-order antibunching [$g^{(3)}(\tau_1,\tau_2)$] (b,f), and fourth-order antibunching [$g^{(4)}(\tau_1,\tau_2,\tau_3)$] (c,d and g,h) at zero delay time presented as 2D cross sections at $\tau_3 = 0$ (c,g) and at $\tau_3 = 1 \cdot t_{Pulse}$ (d,h) delay times, of an example single NPL from the small-sized sample (a-d) and a second one from the large-sized sample (e-h). The experimental parameters of the measurements: excitation power of the laser: ~220 nW; repetition rate of the laser: 5 MHz; measurement time: 180 s.

## Supporting Information section S3: Saturation experiments

Photon correlation measurements from single nanocrystals are typically independent of the excitation power below saturation intensities. To assure that the effect of saturation does not play a role in our results we have measured the PL of single NPLs *versus* the excitation power and determined a region of linear response.

To examine saturation of single NPLs we use the same microscope setup described in the main text with a focused excitation beam (FWHM of ~400 nm). In a saturation experiment the power of the laser was gradually changed with time in cycles of 30 seconds following a triangular wave pattern, at a relatively low laser repetition rate (1 MHz). A typical PL trace from a single CdSe/CdS NPL from the medium sized sample is shown in figure S5a (blue curve) alongside the number of laser photons per pulse (orange line). Generally, a flat-top shape of the PL signal curve indicates nanocrystal saturation, and similar heights of all peaks assure that photobleaching has not occurred. To clearly present these results, figure S5b shows a 2D histogram of fluorescence signal and excitation power.[1] In the range of excitation densities used here, substantial saturation does not occur. Note that the maximal pulse energy here is $\sim 2.5 \cdot 10^{-13}$ J. In order to avoid the effects of saturation, our following experiments did not exceed $5 \cdot 10^{-14}$ J energy per pulse while keeping the same excitation spot profile.

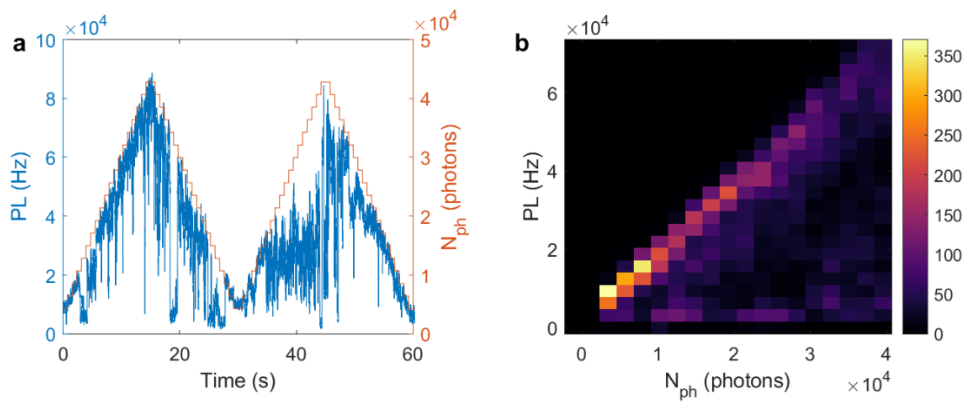

**Figure S5.** (a) Fluorescence trace of a single CdSe/CdS NPL from the medium sized sample (blue line) and the number of laser photon passing through the sample plane per pulse (orange line). For clarity, we present only a third of the time trace used here. (b) A 2D histogram of fluorescence signal and number of excitation laser photons. We attribute the diagonal line to 'on' state emission; its linearity suggests that the maximal power is still below the saturation power.

# Supporting Information section S4: Calculation of the average number of absorbed photons per pulse

In order to corroborate the results of saturation measurements, this section provides an estimate for the average number of generated excitons per laser pulse in a NPL positioned at the focal spot of the laser in our experiment.

In general, generation of an exciton *via* photon absorption occurs both in the core and the shell of a NPL. However, since the 3ML CdS lowest transition is above 3 eV whereas the excitation photon's energy here is ~2.6 eV, we assume that photons are almost entirely absorbed in the CdSe core. Indeed the supporting information of reference shows that the addition of a CdS shell increases absorption of the core only by ~20%, probably due to concentration of the electric field.[2]

To estimate the absorption cross-section of the NPLs synthesized here we use the data published by Yeltik *et al.* which show that the absorption cross-section of a 5ML NPL with an area of $A = 665$ nm$^2$ at 2.6 eV is $\sigma_{Abs} \cong 5 \cdot 10^{-14}$ cm$^2$.[3] Assuming a linear relation between the absorption cross-section and the area of a NPL[3] we estimate $\sigma_{Abs}^S = 5 \cdot 10^{-15}$ cm$^2$, $\sigma_{Abs}^M = 2 \cdot 10^{-14}$ cm$^2$ and $\sigma_{Abs}^L = 4 \cdot 10^{-14}$ cm$^2$ for the small, medium, and large area NPL samples used in this work.

Assuming a Gaussian profile for the excitation beam

$$I(r) = A \cdot \exp\left(-\frac{r^2}{2\Delta^2}\right), \tag{S7}$$

we obtain an expression for the average number of photons absorbed per pulse in a single NPL for a laser with a power $P_L$, a pulse repetition rate $f_{pulse}$, and energy per photon $E_{ph}$

$$\langle n \rangle^{(i)} = \frac{P_L}{f_{pulse} \cdot E_{ph}} \cdot \frac{\sigma_{Abs}^i}{2\pi\Delta^2} \tag{S8}$$

For the parameters of this experiment $P_L = 90$ nW, $f_{pulse} = 5 \cdot 10^6$ Hz, $\Delta \sim 250$ nm and $E_{ph} \cong 4.1 \cdot 10^{-19}$ J, the average exciton population immediately after the pulse is $\langle n \rangle^{(S)} = 0.04$, $\langle n \rangle^{(M)} = 0.2$ and $\langle n \rangle^{(L)} = 0.39$ for the small, medium and large area samples.

While the value of $\langle n \rangle^{(L)}$ approaches the exciton population saturation, we note that even in this case the population of biexcitons is 5 times smaller than that of the single excitons and the corrections for $g^{(2)}(0)$ are very small.[4]

## Figure S6: Typical blinking curve for a single NPL

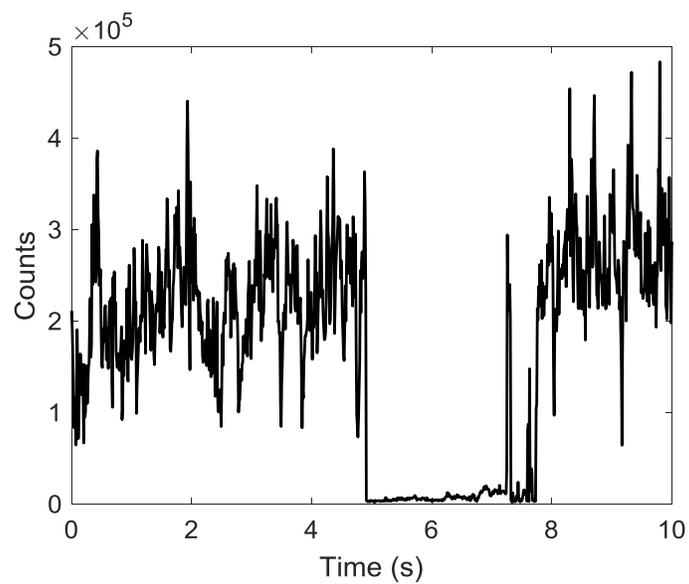

**Figure S6.** Blinking trace of single medium-NPL showing that the intensity of the 'off state' is approaching zero.

# Figure S7: Second-order correlation *versus* time gating of photon detections

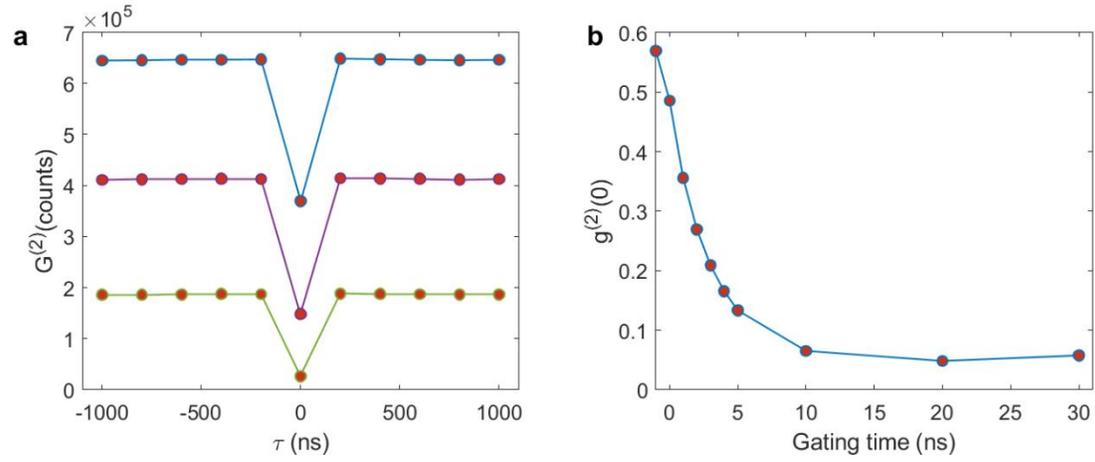

**Figure S7.** In order to test whether a measurement is taken from a single particle we use a time gating test. We apply a filter in post-processing, discarding of detections that arrive before the gating time. (a) $G^{(2)}$ curve for three representative gating times (out of ten); -1 ns delay (blue), +1 ns delay (purple), and +5 ns delay (green) relative to the maximum of the lifetime. With increasing gating time we observe a decrease in the $g^{(2)}(0)$ values. (b) Normalized second-order correlation values at zero delay ($g^{(2)}(0)$) for increasing gating times showing a drastic decrease. The $g^{(2)}(0)$ values of each curve in (a) translates into a point in (b).

# Figure S8: Distribution of $g^{(2)}(0)$ for single particle measurements from three different samples

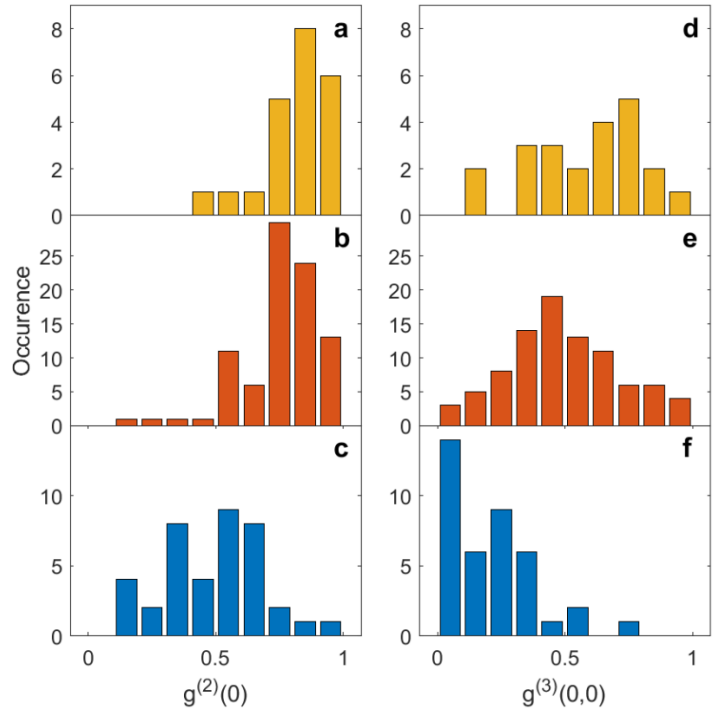

**Figure S8.** Histograms of Second and third order correlation values at zero delay for an ensemble of measured particles from the large (a, d), medium (b, e) and small (c, f) NPL samples, respectively. Since the SNR of $g^{(4)}$ values is relatively low, we did not present their histograms here.

**Table S1.** Average values (standard deviations) of the correlation functions at zero delay times.

| NPL size | $\langle g^{(2)}(0) \rangle$ | $\langle g^{(3)}(0,0) \rangle$ | $\langle g^{(4)}(0,0,0) \rangle$ |
|----------|------------------------------|--------------------------------|----------------------------------|
| Small    | 0.5 (0.2)                    | 0.21 (0.17)                    | 0.07 (0.1)                       |
| Medium   | 0.75 (0.16)                  | 0.5 (0.24)                     | 0.28 (0.32)                      |
| Large    | 0.82 (0.14)                  | 0.57 (0.22)                    | 0.38 (0.37)                      |

# Supporting Information section S5: Modeling photon correlation for bound excitons

In the following, we re-formulate a model for the recombination kinetics of multi-excitonic states through radiative recombination of isolated excitons and Auger recombination of exciton-exciton collisions. We apply the model to calculate the values of second-, third- and fourth-order correlations.

In more detail, the model assumes that charge carriers exist as bound electron-hole pairs. Moreover, we disregard the non-radiative recombination of single excitons such that only radiative recombination and biexciton Auger recombination count as decay pathways.

The second-order correlation at zero delay time in the case of a biexciton can be defined as:

$$g^{(2)}(0) = \frac{2k_{1x}^{rad}}{2k_{1x}^{rad} + k_{2x}^{Aug}} \tag{S9}$$

The pre-factor for each radiative decay rate in the above equation is a result of the multiple pathways for each process in the multi-excited state. For example, the term $2k_{1x}^{rad}$ represents two times radiative decay rate of a single exciton in a biexciton emission process. For a process that requires m excitons in an n exciton state the multiplicity of pathways is $\binom{n}{m}$ ("n choose m").

Similarly, the third-order correlation at zero delay time in the case of a triexciton can be calculated as:

$$g^{(3)}(0,0) = \frac{3k_{1x}^{rad}}{3k_{1x}^{rad} + 3k_{2x}^{Aug}} \cdot g^{(2)}(0) = \frac{[g^{(2)}(0)]^2}{2 - g^{(2)}(0)} \qquad \text{(S10)}$$

The last equality suggests that under this model $g^{(3)}(0,0)$ can be represented as a function of $g^{(2)}(0)$ alone. As was explained above, for an Auger process of a triexciton the multiplicity of pathways is $\binom{3}{2} = \frac{3!}{2!1!} = 3$.

Note that for simplicity we do not consider here the single exciton non-radiative recombination process (through e.g. charge trapping). Incorporating this option in the model yields more cumbersome mathematical expressions, without varying the results for the dependence of third-order correlation on the magnitude of the second-order correlation (Eq. S10S10).

Using the same logic, the fourth-order correlation at zero delay time in the case of quad-exciton can be defined as:

$$\begin{aligned} g^{(4)}(0,0,0) &= \frac{4k_{1x}^{rad}}{4k_{1x}^{rad} + 6k_{2x}^{Aug}} \cdot g^{(3)}(0,0) \\ &= \frac{[g^{(2)}(0)]^3}{[3 - 2g^{(2)}(0)] \cdot [2 - g^{(2)}(0)]} \end{aligned} \qquad \text{(S11)}$$

and can also be represented as a function of $g^{(2)}(0)$.

The relations of $g^{(3)}(0,0)$ *versus* $\left[g^{(2)}(0)\right]^2$ and $g^{(4)}(0,0,0)$ *versus* $\left[g^{(2)}(0)\right]^3$ are plotted in the main text, figure 4a,b respectively. Here, the multiplicity of pathways for Auger process is calculated as follows; $\binom{4}{2} = 6$.

We have also considered a model for the charges being free rather than bound, giving identical results. However, it is reasonable to assume that most charge carriers act as bound excitons because of high binding energies that result from the nanoplatelets' geometry.[5]

# Supporting Information section S6: Phenomenological model of photon correlations for bound excitons

As mentioned in the discussion regarding figure 4 of the main text, the presented results deviate from the models presented in Supporting Information sections 5 and 6, suggesting that a multi-body interaction is necessary to complete the picture.

In an attempt to quantitatively assess the magnitude of multi-body effect, we suggest here a modified model to account for the relation between second- and third-order antibunching in the measurements of single NPLs, shown in figure 4 of the main text. The results were compared to the simplified binary collision model described above, but most of the results present small deviation to lower $g^{(3)}(0,0)$ values. In order to correct for that, the results were also fit to a phenomenological model, which adds a contribution of 3-body decay rate term ($k_{3B}$) in the Auger decay of triexciton.

$$g^{(3)}(0,0) = \frac{3k_{1x}^{rad}}{3k_{1x}^{rad} + 3k_{2x}^{Aug} + k_{3B}} \cdot g^{(2)}(0) \qquad (S12)$$

Re-arranging equation S9 we get

$$k_{1x}^{rad} = \frac{1}{2} \cdot \frac{g^{(2)}(0)}{1 - g^{(2)}(0)} \cdot k_{2x}^{Aug} \qquad (S13)$$

Finally, using the expression for $k_{1x}^{rad}$ from equation **Error! Reference source not found.** in equation **Error! Reference source not found.** we obtain

$$g^{(3)}(0,0) = \frac{\left[g^{(2)}(0)\right]^2}{(2+\alpha)-(1+\alpha)\cdot g^{(2)}(0)}, \tag{S14}$$

where $\alpha = \frac{3}{2}\frac{k_{3B}}{k_{2x}^{Aug}}$.

As mentioned in the main text, fitting the data to this model yielded a $\alpha = 0.19 \pm 0.05$.